\documentclass[a4paper]{jpconf}
\usepackage{graphicx}
\begin{document}
\title{Deeply virtual Compton Scattering off $^4$He}

\author{Sara Fucini, Sergio Scopetta}

\address{Dipartimento di Fisica e Geologia, University of Perugia and INFN, Perugia, 
\\
I-06100 Perugia, Italy}
\ead{sara.fucini@pg.infn.it,
sergio.scopetta@pg.infn.it}

\author{Michele Viviani}

\address{INFN-Pisa, I-06100 Pisa, Italy}

\ead{michele.viviani@pi.infn.it}

\begin{abstract}

Deeply virtual Compton scattering is a fascinating process which can provide a tomographic view of nuclei and bound nucleons.
The first experimental results for $^4$He targets, recently released at Jefferson Lab, have been
analyzed here in a rigorous Impulse Approximation scenario. For both the
coherent and incoherent channels of the process, the main experimental observables have been written in terms of state-of-the-art models of the nuclear spectral function and of the
parton structure of the bound proton. A good overall agreement with the
data is obtained. The calculation shows that a comparison of our
conventional results
 with future precise data can expose novel quark and gluon effects in nuclei.

\end{abstract}

\section{Introduction}
It is nowadays
clear that inclusive Deep Inelastic Scattering 
measurements
do not allow a
quantitative understanding of the origin of the EMC effect \cite{aubert}, i.e. the nuclear medium modification to the parton structure of the bound nucleon. 
Nevertheless, a new generation of semi-inclusive and exclusive experiments, performed in particular at Jefferson Lab (JLab), are expected to give new insights into the problem
\cite{duprescopetta,cloettrento}.
 A powerful tool in this sense is deeply Virtual Compton Scattering (DVCS).
In DVCS, the inner parton content of the target is parametrized through non-perturbative functions, the so-called generalized parton distributions (GPDs), which provide a wealth of novel information (for an exhaustive report, see, e.g., Ref. \cite{diehlgpd}). 
The one directly linked to this talk is the possiblity to obtain a parton tomography
of the target \cite{tomografia}.
In a nucleus, such a process can occur in two different channels: the coherent one, where the nucleus remains intact  and the tomography of the whole nucleus can be accessed, and the incoherent one,  where the  nucleus breaks up,
one nucleon is detected and its structure can be studied. As a target, $^4$He is very convenient, being the lightest system showing the dynamical features of a typical atomic nucleus. Moreover, it is scalar and isoscalar and its description in terms of GPDs is easy.
Recently, DVCS data for this target have become available at JLab where the coherent and incoherent channels have been successfully disentangled,
for the first time \cite{hattawycoerente,hattawyincoerente}. 

A rigorous theoretical description of the process, whose results could be compared with the data in a conclusive way, requires a proper evaluation of  conventional nuclear physics effects in terms of wave functions corresponding to realistic nucleon-nucleon potentials. This kind of realistic calculations, although very challenging, are possible for a few-body system (e.g. see Ref. \cite{cano} for $^2$H and Ref. \cite{Scopetta} for $^3$He) as the target under scrutiny. Previous calculations for $^4$He have been performed long time ago
\cite{liuticoerente,guzey}, in some cases in kinematical regions different from those probed at JLab.
In this talk, a review of our main results obtained from the study of the handbag contribution to both DVCS channels, in impulse approximation, is presented.
\begin{figure}[t]
    \centering
    \includegraphics[scale=0.40]{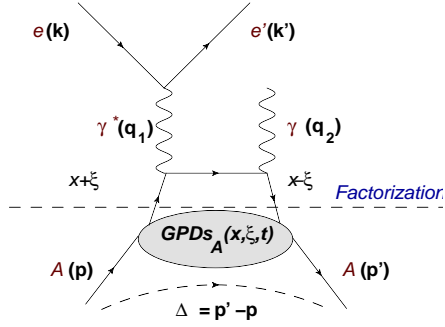}
    \caption{The handbag approximation to the coherent DVCS of $^4$He.}
    \label{figcoh}
\end{figure}
\section{DVCS formalism}
In this section, the general formalism for both DVCS channels, whose handbag approximation will be studied in Impulse Approximation (IA), is presented. In this scenario, we assume that the process
occurs off one quark in one nucleon in $^4$He,
that only nucleonic degrees of freedom are considered, and that
further possible rescattering of the struck proton
with the remnant systems is not relevant. 
As reference frame, we choose the target at rest, with an azimuthal angle $\phi$
between the electron scattering
plane and the hadronic production plane.
The process can be described in terms of four independent variables, usually chosen as $x_B = Q^2/(2 M \nu)$, $Q^2 =-q_1^2 = -(k-k')^2$, $\Delta^2 =(p'-p)^2 = (q_1-q_2)^2$ and $\phi$. 
In such a process, if the initial photon
virtuality $Q^2$ is much larger than the momentum transferred
to the hadronic system with initial, the factorization property allows to distinguish the hard vertex, which can be studied perturbatively, from the soft part, given by the blob in Figs \ref{figcoh} and \ref{incodvcs}, that is parametrized in
terms of GPDs. Besides $Q^2$ and $t$, GPDs are also a function of the so-called skewness $\xi = -\frac{\Delta^+}{P^+}$ i.e., the difference in plus momentum fraction between the initial
and the final states, and on $x$, the average plus momentum
fraction of the struck parton with respect to the
total momentum, not experimentally accessible.
Because of this latter dependence, GPDs cannot be directly measured. For this reason, Compton Form Factors (CFFs), where GPDs ($H_q)$ are hidden, are defined in the following way ($e_q$ being the quark electric charge):
\begin{equation}\label{imeq}
  \Im m {\cal H}(\xi,t)=\sum_{q} e_q^2(H_q(\xi,\xi,\Delta^2)- H_q(-\xi,\xi,\Delta^2))\,,
\end{equation}
\begin{equation}\label{re}
	\Re e {\cal H}(\xi,t)= \Pr \sum_{q}e_q^2 \int_{0}^{1}\bigg(\frac{1}{\xi-x}-\frac{1}{\xi+x}\bigg)(H_q(x,\xi,t)-H_q(-x,\xi,t))
	\end{equation}
The experimental observable which gives access to these quantities is the beam spin asymmetry (BSA), that for the target under scrutiny is given by

\begin{equation}\label{alu}
    A_{LU}= \frac{d\sigma^+ - d\sigma^-}{d \sigma^+ + d\sigma^-}\,,
\end{equation}
where the differential cross section for the different beam polarization ($\pm$) appears.
A realistic
calculation of conventional effects for the BSA corresponds to a plane wave impulse
approximation analysis, presented in the following. 

\section{Coherent DVCS channel} 
The most general coherent DVCS process $A(e,e'\gamma)A$ allows to study the partonic structure of the recoiling whole nucleus $A$ through the formalism of GPDs. In the IA scenario presented above, a workable expression for
$H_q^{^4He}(x, \xi,\Delta^2)$, the GPD of the quark of flavor q in the $^4$He nucleus,
is obtained as a convolution between the GPDs $H_q^N$of the quark of flavor $q$ in the bound
nucleon N and the off-diagonal light-cone momentum distribution
of N in $^4$He and reads

\begin{equation}
H_q^{^4He}(x,\xi, \Delta^2)= \sum_N \int_{|x|}^1 { dz \over z } 
	h_N^{^4He}(z,\xi,\Delta^2)
H_q^N\bigg(\frac{x}{\zeta},\frac{\xi}{\zeta},\Delta^2\bigg).
\label{gpd}
\end{equation}
The light cone momentum distribution in the previous equation is defined as \begin{equation}
h_N^{^4He}(z,\Delta^2,\xi)= 
\int dE \, \int d \vec p \,
P^{^4He}_N(\vec p, \vec p + \vec \Delta, E) \delta \bigg(z - \frac{\bar p^+}{ \bar P^+}\bigg)\,,
\end{equation}
where the off diagonal spectral function $P_N^{^4He}(\vec p, \vec p + \vec \Delta,E)$ governs the size and relevance of nuclear effects. It represents the probability
amplitude to have a nucleon leaving the nucleus with momentum $\vec p$ and leaving
the recoiling system with an excitation energy
$E^*=E-|E_A|+|E_{A-1}|$, with $|E_A|$ and $|E_{A-1}|$ the nuclear binding energies,
and going back to the nucleus with a momentum
transfer $\vec \Delta$.
\begin{figure}
\center
    \hspace{-0.5cm}
    \includegraphics[angle=270,scale=0.22]{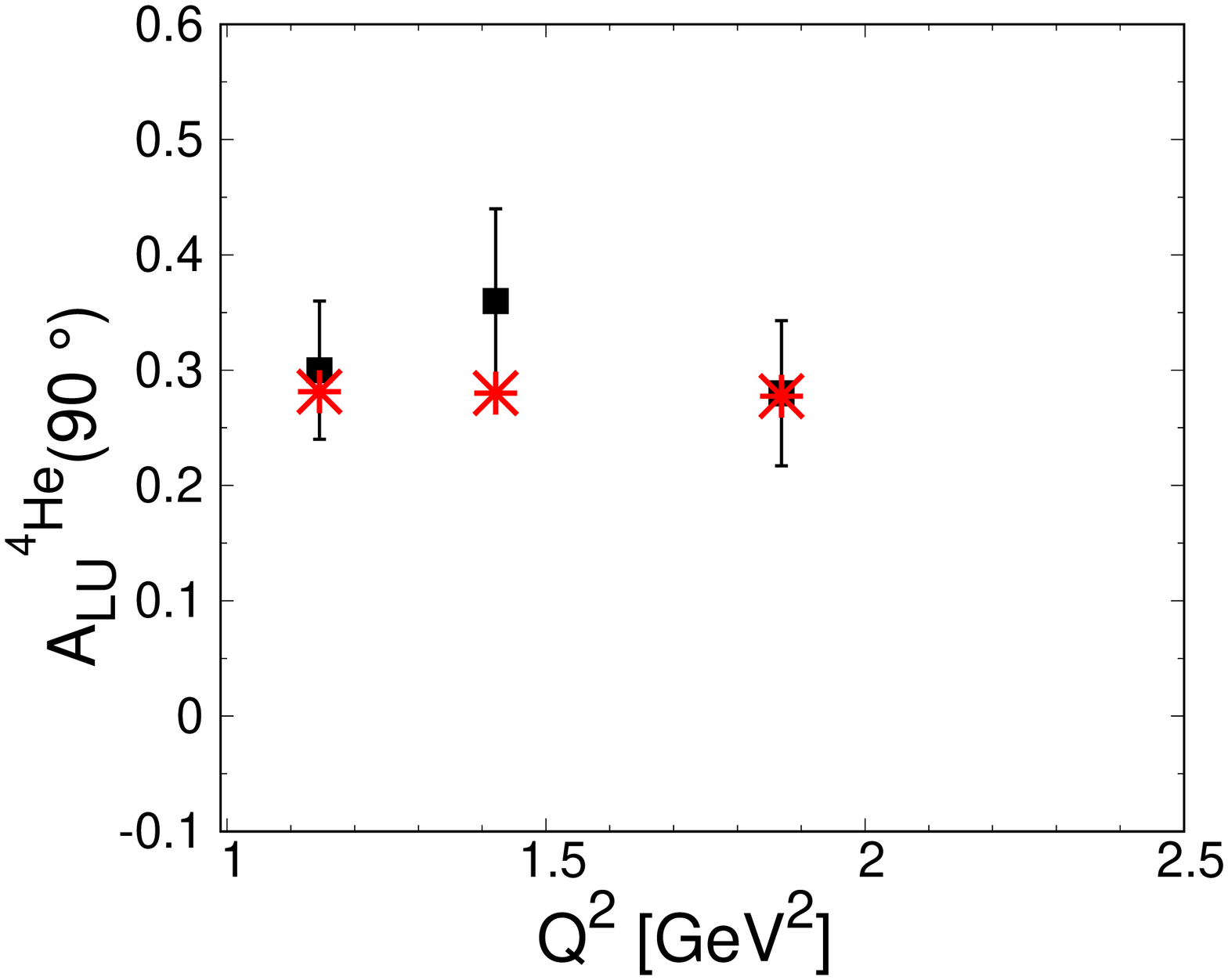}
    \hspace{-0.5cm}
    \includegraphics[angle=270,scale=0.22]{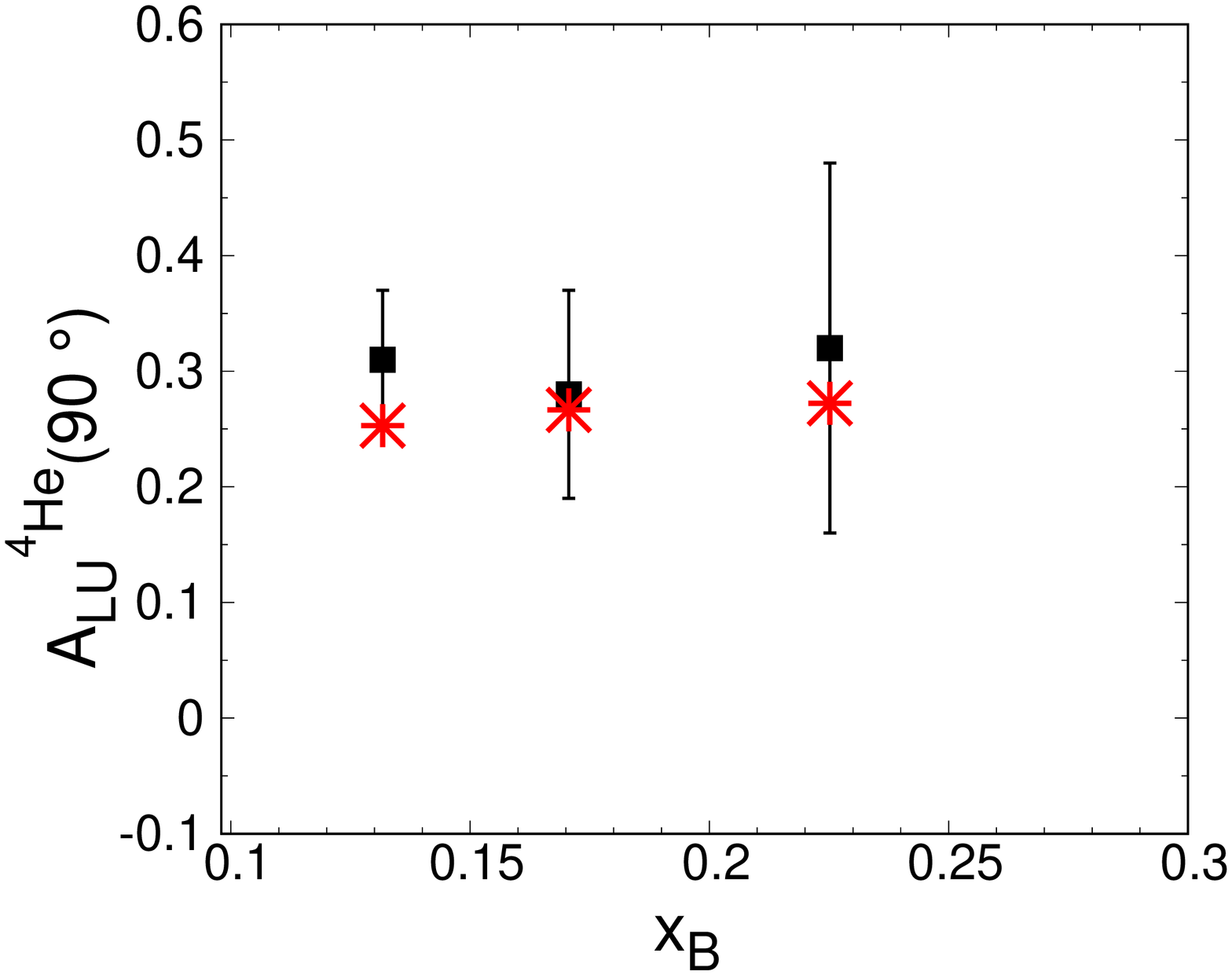}
    \hspace{-0.5cm}
    \includegraphics[angle=270,scale=0.22]{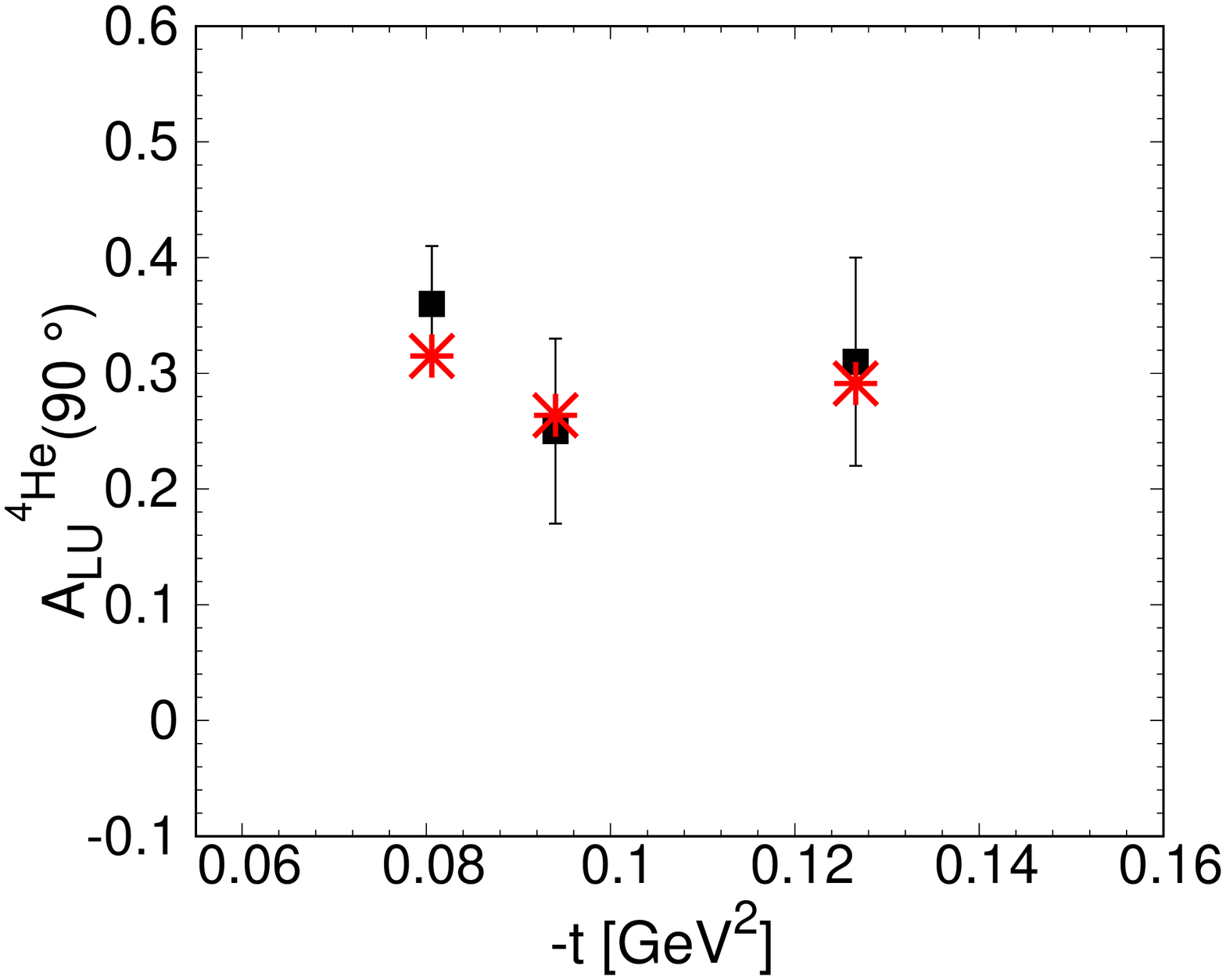}
    \hspace{-0.5cm}
    \caption{$^4$He azimuthal beam-spin asymmetry
$A_{LU}(\phi=90^o)$: results of Ref. \cite{nostrocoerente} (red stars) compared
with data (black squares) \cite{hattawycoerente}. From left to right, the quantity
is shown in the experimental $Q^2$, $x_B$ and $t=\Delta^2$ bins, respectively.}
    \label{alucoerente}
\end{figure}
The full realistic evaluation of $P_N^{^4He}$ requires an exact
description  of all the
$^4$He spectrum, including three-body scattering states; for this reason, it represents a challenging, presently unsolved few body problem. So, while the complete evaluation of this object has just begun, as an
intermediate step in the present calculation a
model of the nuclear non-diagonal spectral function \cite{vivianikievsky},
based on the momentum distribution corresponding
to the Av18 NN interaction Ref.\cite{av18}
and including 3-body forces \cite{3bf}, has been used when excited
3- and 4- body states
are considered.
For the ground state, exact wave functions of 3- and 4-body systems,
evaluated along the scheme of Ref. \cite{rosati},
have been used.
Concerning the nucleonic GPD appearing in Eq. (\ref{gpd}), the well known GPD model of Ref. \cite{golkroll} has been used.
With these ingredients at hand, as an encouraging check, typical results are found, in the proper limits,
for the nuclear charge form factor and for nuclear parton
distributions. In this way, our model for $H_q^{^4He}$ allowed us to have a numerical evaluation of Eqs. (\ref{imeq}) and (\ref{re}), which define
quantities also appearing in the explicit form of the BSA of the coherent DVCS channel that reads:
\begin{equation}
A_{LU}(\phi) = 
\frac{\alpha_{0}(\phi) \, \Im m(\mathcal{H}_{A})}
{\alpha_{1}(\phi) + \alpha_{2}(\phi) \, \Re e(\mathcal{H}_{A}) 
+ \alpha_{3}(\phi) 
\bigg( \Re e(\mathcal{H}_{A})^{2} + \Im m(\mathcal{H}_{A})^{2} \bigg)} \,.
\end{equation}

Here above, $\alpha_i(\phi)$ are kinematical coefficients  defined in Ref. \cite{belmullernucleo}.
As shown in Fig. \ref{alucoerente}, a very good agreement is
found with the data
\cite{nostrocoerente}. One can conclude that a careful
analysis of the reaction mechanism in terms of basic
conventional ingredients is successful and that the
present experimental accuracy does not require the
use of exotic arguments, such as dynamical off-shellness.

\section{Incoherent DVCS channel}
\begin{figure}
    \centering
    \includegraphics[scale=0.35]{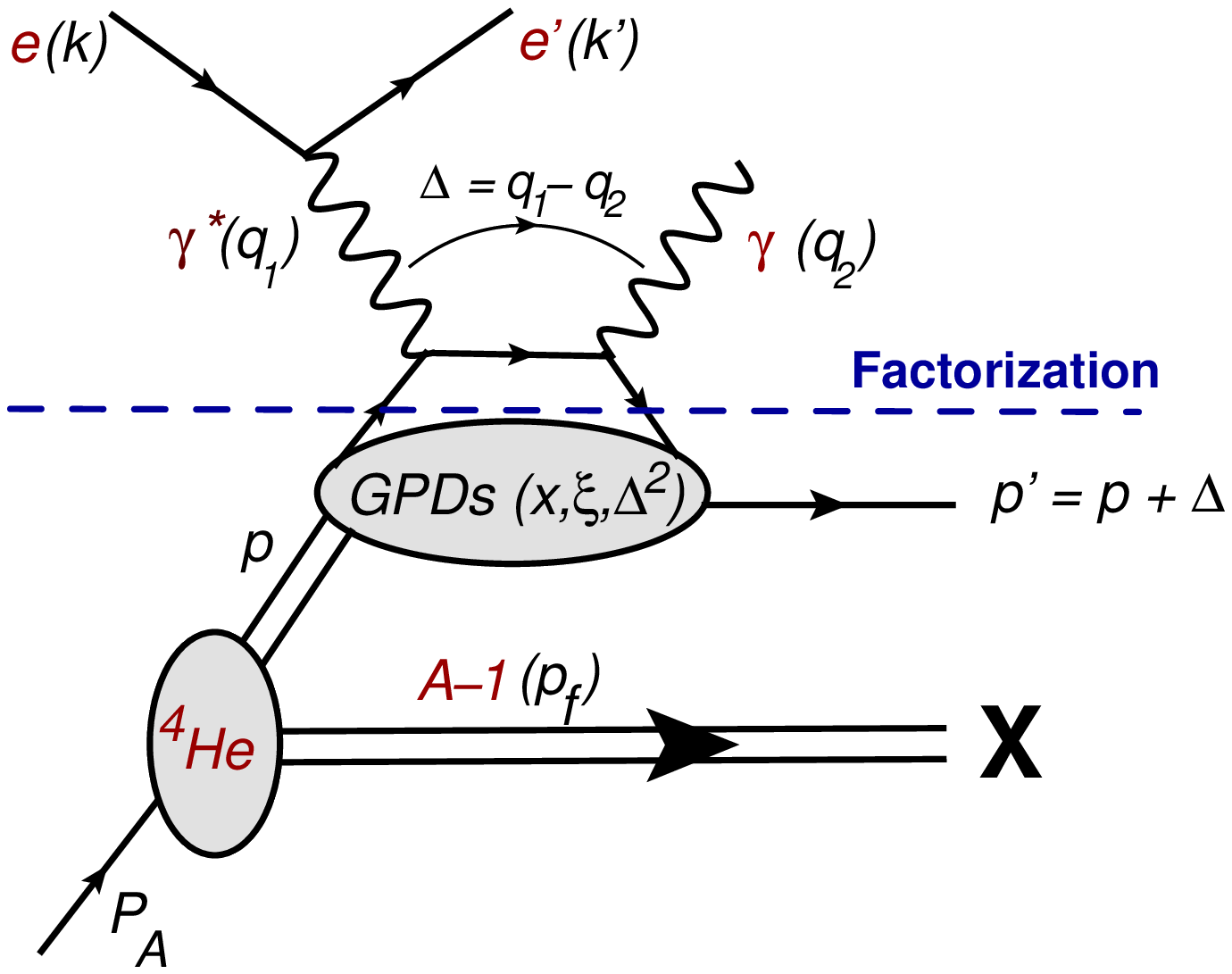}
    \hspace{0.5cm}
    \includegraphics[scale=0.35]{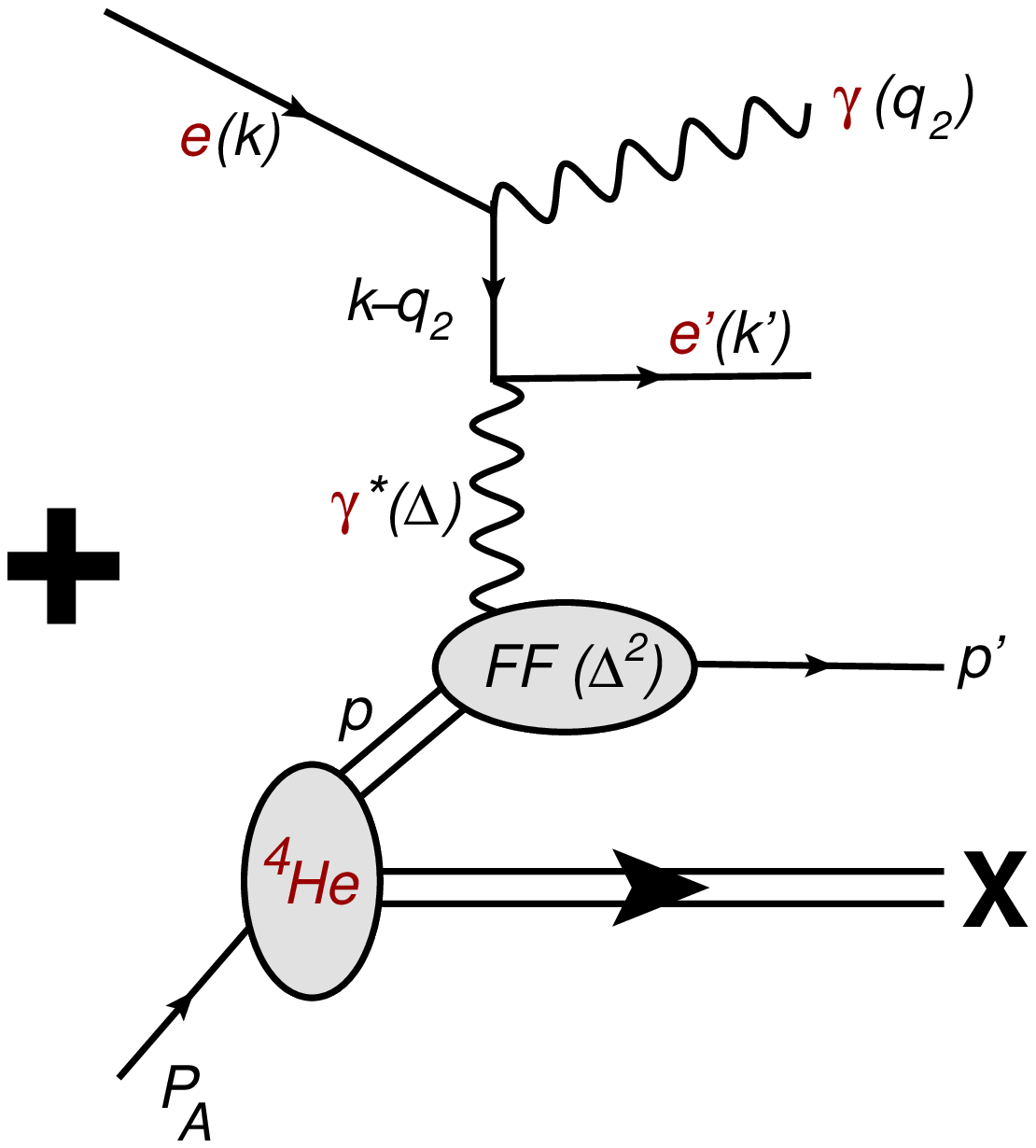}
    \hspace{0.5cm}
     \includegraphics[scale=0.35]{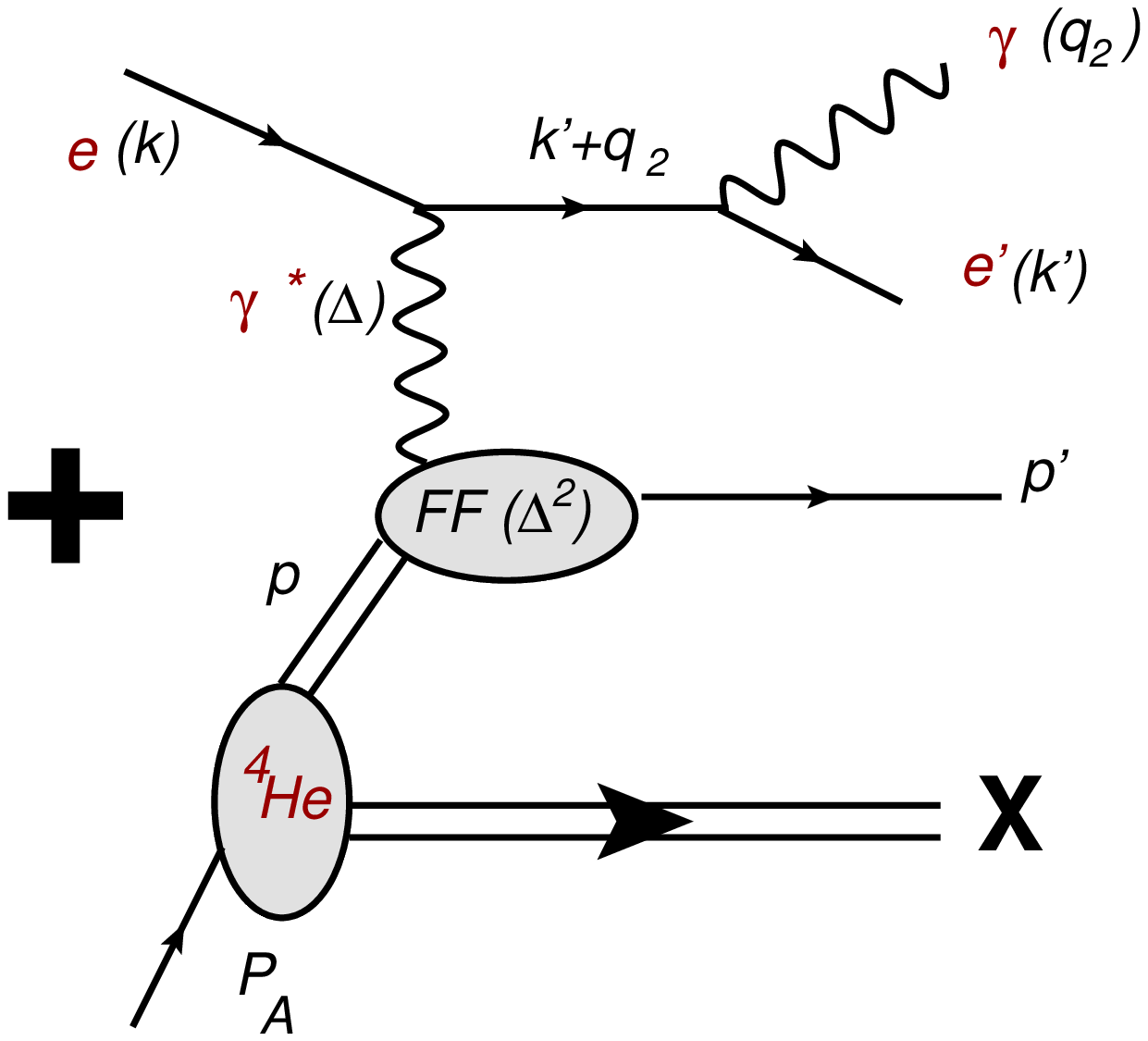}
    \caption{Incoherent DVCS off $^4$He in IA. To the left, pure DVCS contribution; to the right the two Bethe Heitler terms.}
    \label{incodvcs}
\end{figure}
\begin{figure}
    \centering
    \hspace{-0.5cm}
    \includegraphics[scale=0.40]{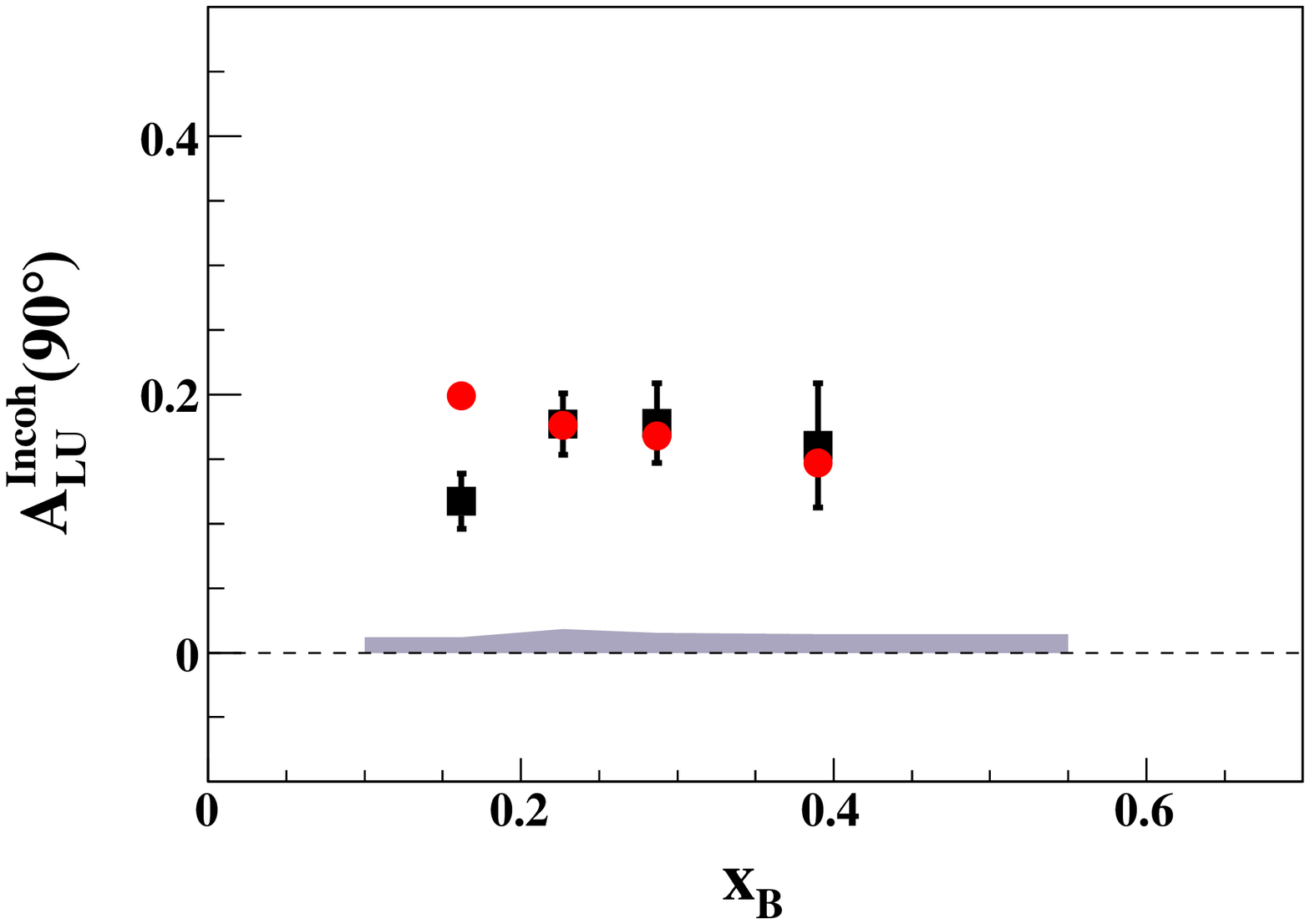}
    \caption{
Azimuthal beam-spin asymmetry for the proton in
$^4$He, $A_{LU}^{Incoh}$,
Eq. (\ref{aluratio}),
for $\phi = 90^o$: results of this approach \cite{nostroincoerente}(red dots) compared with data
(black squares) \cite{hattawyincoerente}.}
\label{4}
     \hspace{-0.5cm}
    \includegraphics[scale=0.40]{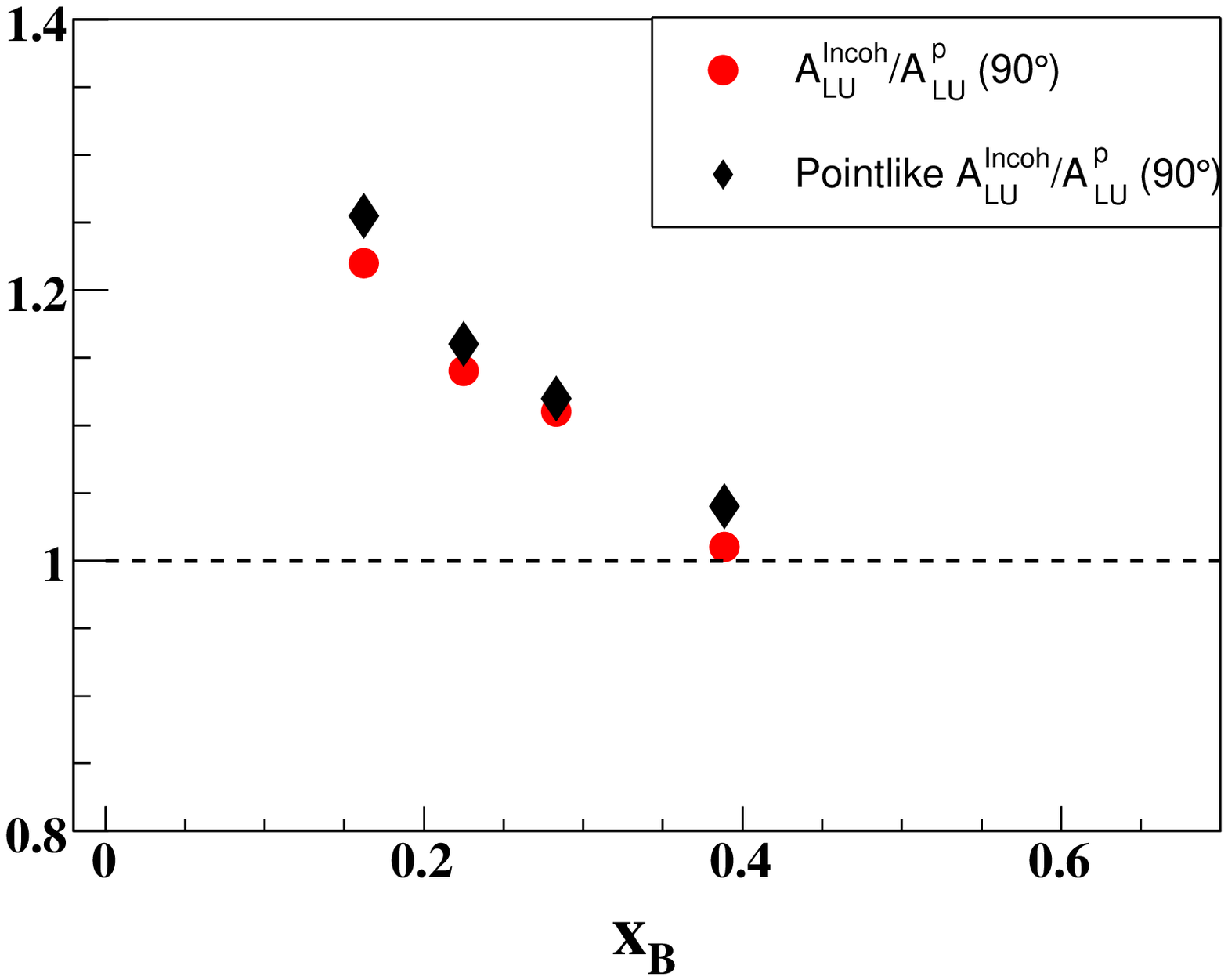}
    \caption{The ratio 
$A_{LU}^{Incoh} / { A_{LU}^p }$,
Eq. (\ref{aluration}) (red dots),
compared to the result obtained with
pointlike protons (black diamonds).}
\label{aluxbinco}
\end{figure}

In the process $A(e,e'\gamma p)X$  depicted in Fig. \ref{incodvcs}, the parton structure of the bound proton can be accessed. In order to have a complete evaluation 
of Eq. (\ref{alu}), the cross-section for a DVCS process occurring off a bound moving proton in $^4$He is required.  Working within an IA approach, we account for the pure kinematical off-shellness of the initial bound proton obtaining a convolution formula for the cross sections differential in the experimental variables: 
\begin{equation}
d \sigma^\pm \equiv  \frac{d\sigma^\pm_{Inc} }{d{x}_B dQ^2 d{\Delta}^2 d\phi}= \int_{exp} dE \, d{\vec p} \,P^{^4He}(\vec {p},E)
|\mathcal{A}^{\pm}({\vec p}, E ,K)|^2
g(\vec{p},E,K) \nonumber \,,
\end{equation}
where $K$ is the set of kinematical variables $\{x_B,Q^2,t,\phi\}$.
The intervals of these variables
probed in the experiment
select the relevant
part
of the
diagonal spectral function $P_N^{^4He}(\vec p, E)$,
which has therefore to be integrated in the range $exp$.
The quantity
$ g({p},{p}_N,K)$ is a complicated function arising from the integration over the phase space and including also the flux factor ${p\cdot k}/({p_0 \,|\vec  k |})$. 
In the above equation, the squared amplitude includes three different terms, i.e  $\mathcal{A}^2= T_{DVCS}^2+T_{BH}^2+\mathcal{I}_{DVCS-BH}$ as shown in Fig. \ref{incodvcs} and each contribution has to be evaluated for an initially moving proton.
Our amplitudes generalizes the ones obtained for a proton at rest in Ref. \cite{belitskynucleon} and the main assumptions done are  summarized in Ref. \cite{nostroincoerente}. 
Since in the kinematical region of interest at Jlab the BH part is dominating, the key partonic insights are all hidden in the interference DVCS-BH term entering in the BSA in the following way
\begin{equation}\label{aluratio}
A_{LU}^{Incoh} = \frac{
\int_{exp} dE \, d \vec p \, 
{ {P^{^4He}(\vec p, E )}} 
\, g(\vec p,E,K)\, 
{{\mathcal{I}_{DVCS-BH}}} 
}
{
\int_{exp} dE \, d \vec p \, {{P^{^4He} (\vec p, E )}} 
\, g(\vec p, E,K)\,
{{T_{BH}^2}}
} .
\end{equation}
Since our ultimate goal is to have a comparison with the experimental data, we exploit the azimuthal dependence of Eq. (\ref{aluratio}) decomposing in $\phi$ harmonics the interference and the BH part. 
All the information about the parton content of the bound proton is encapsulated in the imaginary part of CFF, that accounts for the modification at structure level through the rescaling of the skewness, that depends explicitly on the 4-momentum components of the initial proton.  In the present calculation we considered only the dominating contribution given by the $H_q(x,\xi',t)$ GPD, for which use of the GK model has been made \cite{golkroll}. The results are depicted in Fig. \ref{4} \cite{nostroincoerente}. As expected, the agreement with experimental data is good except the region of lowest $Q^2$, corresponding to the first $x_B$ bin. In this region, in facts, the impulse approximation is not supposed to work well, since final state interaction effects,
neglected in IA, could be 
sizable.
In order to have an idea about how the nuclear effects affect the results obtained, i.e. if they are related to some medium modification of the inner parton structure 
described by the GPD, we considered the ratio between the BSA for a bound nucleon, given by Eq. (\ref{aluratio}) and that for a free proton,
given in our scheme by the
GK model: 
\begin{equation}
\frac{ A_{LU}^{Incoh} } { A_{LU}^p } 
\propto \frac{ \mathcal{I}_{DVCS-BH}^{^4He} } { \mathcal{I}_{DVCS-BH}^{p} }
\frac{ T_{BH}^{2\, \, p} } { T_{BH}^{2\, \, ^4He} }= \frac
{(nucl.eff.)_{Int}}{(nucl.eff.)_{BH}}\,.
\label{aluration}
\end{equation}
The above quantity
 is proportional to the ratio
of the nuclear effects 
on the $BH$ and $DVCS$ interference
$\mathcal{I}_{DVCS-BH}$
to
the nuclear effects on the $BH$ cross section.
If the nuclear dynamics modifies $\mathcal{I}_{DVCS-BH}$ and the BH cross sections
in a different way, the effect
can be big even if the parton structure
of the bound proton does not change appreciably.
Indeed, this is what happens:
considering the same ratio for pointlike protons, the big observed effect is still present, as we can see from Fig. \ref{aluxbinco}.

\section{Conclusions}
We can conclude
that for both channels, given the present experimental accuracy, the description of the data does not require the
use of exotic arguments, such as dynamical off shellness.
Nevertheless, a serious benchmark calculation in the kinematics
of the next generation of precise measurements at high
luminosity \cite{armstrong} will require an improved treatment of both the
nucleonic and the nuclear parts of the evaluation. The latter
task includes the realistic computation of a one-body non diagonal (for the coherent channel) and diagonal (for the incoherent channel)
spectral function of $^4$He. Work is in progress towards this
challenging direction. In the meantime, the straightforward
approach proposed here can be used as a workable framework
for the planning of future measurements.


\section*{References}
\begin{thebibliography}{50}
\bibitem{aubert}
Aubert J J [European Muon Collaboration] 1983 {\it Phys.\ Lett.\ } {\bf 123B}  275

\bibitem{duprescopetta} 
Dupré R and Scopetta S 2016 {\it Eur. Phys. J.} A {\bf 52} 159

\bibitem{cloettrento} Cloët I C {\it et al.} 2019 {\it J.\ Phys.} G {\bf 46} 093001
 
\bibitem{diehlgpd} Diehl M 2003 {\it Phys. Rept}\  {\bf 388} 41;
  Belitsky A V  and Radyushkin A V 2005
  {\it Phys.\ Rept.}\  {\bf 418} 1
  
\bibitem{tomografia} Burkardt M 2000 {\it Phys.\ Rev.} D {\bf 62} 071503;
Erratum:  2002 [{\it Phys.\ Rev.}\ D {\bf 66}, 119903]
 
\bibitem{hattawycoerente}
Hattawy M {\it et al.}, CLAS collaboration 2017 {\it  Phys. Rev. Lett.} {\bf 119} 202004 

\bibitem{hattawyincoerente}
Hattawy M {\it et al.}, CLAS Collaboration 2019 {\it Phys. Rev. Lett.} {\bf 123} 032502 


\bibitem{cano}
 Berger E R, Cano F and Diehl M and Pire B 2001
{\it  Phys.\ Rev.\ Lett.}\ {\bf 87}  142302

\bibitem{Scopetta} 
Scopetta S 
 {\it  Phys.\ Rev.} 2004 \ C {\bf 70}  015205 
 
\bibitem{liuticoerente} Liuti S and Taneja S K 2005 {\it Phys. Rev.} C {\bf 72} 034902;
2005 {\it Phys.  Rev.} C {\bf 72} 032201

\bibitem{guzey} 
  Guzey V and Strikman M 2003
  {\it Phys.\ Rev.}\ C {\bf 68} 015204;
Guzey V, Thomas A W and Tsushima K 2009
{\it Phys.\ Lett.}\ B {\bf 673}

\bibitem{vivianikievsky}
Viviani M, Kievsky A and Rinat A S 2003 {\it Phys. Rev.} C {\bf 67} 034003 

\bibitem{av18} Wiringa R B, Stoks  V G J and Schiavilla R 1995
  {\it Phys.\ Rev.}\ C {\bf 51} 38 
  
\bibitem{3bf} 
Pudliner B S, Pandharipande V, R Carlson J and Wiringa R B  1995 {\it  Phys.\ Rev.\ Lett. }\  {\bf 74} 4396

\bibitem{rosati} Viviani M, Kievsky A and Rosati S 2005 {\it Phys. Rev.} C{\bf 71} 024006

\bibitem{golkroll} Goloskokov S V and Kroll P 2008 {\it  Eur. Phys. J} C {\bf 53} 367 

\bibitem{belmullernucleo} 
Belitsky A V and Muller D 2009 {\it Phys. Rev.} D {\bf 79} 014017 

\bibitem{nostrocoerente} Fucini S, Scopetta S and Viviani M 2018 {\it Phys. Rev.} C {\bf 98} 015203

\bibitem{belitskynucleon}
Belitsky V A, Mueller D and Kirchner A 2002
{\it Nucl.\ Phys.}\ B {\bf 629}, 323

\bibitem{nostroincoerente} Fucini S, Scopetta S and Viviani M 2019 	arXiv:1909.12261 [nucl-th]













\bibitem{armstrong} 
 Armstrong W  {\it et al.},
  arXiv:1708.00888 [nucl-ex]

\end{thebibliography}
\end{document}